\documentclass[reprint,amsmath,aps,pra,footinbib,notitlepage,superscriptaddress]{revtex4-2}
\usepackage{graphicx,epstopdf,dcolumn,bm,mathtools,siunitx,multirow,hyperref,xcolor,amsfonts,physics,bibunits}

\begin{document}


\title{Universal Quantum Birthmark:\\ Ghost of the quantum past} 

\author{I. Xiaoya}
\affiliation{Department of Physics, Harvard University, Cambridge, Massachusetts 02138, USA}
\author{A. M. Graf}
\affiliation{Harvard John A. Paulson School of Engineering and Applied Sciences,
Harvard, Cambridge, Massachusetts 02138, USA}
\author{E. J. Heller}
\affiliation{Department of Physics, Harvard University, Cambridge, Massachusetts 02138, USA}
\affiliation{Department of Chemistry and Chemical Biology, Harvard University, Cambridge,
Massachusetts 02138, USA}
\author{J. Keski-Rahkonen}
\affiliation{Department of Physics, Harvard University, Cambridge, Massachusetts 02138, USA}
\affiliation{Department of Chemistry and Chemical Biology, Harvard University, Cambridge,
Massachusetts 02138, USA}
\affiliation{Computational Physics Laboratory, Tampere University, P.O. Box 600, FI-33014 Tampere, Finland}

\begin{abstract}
Quantum dynamics retains a permanent and universal memory of its initial conditions, even in systems whose spectra display fully chaotic, random-matrix behavior. This effect, known as the quantum birthmark, appears as an enhancement of the long-time return probability of any non-stationary state compared to the overlap with a typical ergodic state. In this work, we develop the full theoretical foundation for this universal contribution that depends only on the global symmetry class and accessible Hilbert-space dimension, not on the microscopic dynamics. Our findings reveal that quantum evolution preserves an unavoidable, symmetry-controlled imprint of its origin, a quantum effect calling into question classical expectations of ergodicity and the resulting thermalization scenarios.
\end{abstract}

\maketitle

\section{Introduction}
\noindent
The most fundamental tenet of classical chaos is ergodicity, i.e., the long-time tendency of trajectories to “forget’’ their initial conditions and uniformly explore all \emph{a priori} accessible regions of phase space.~\cite{Strogatz_book, Arnold_book} This assumption, for instance, underlies the zfoundations of thermodynamics.~\cite{Gallavotti_book} In quantum systems, however, this classical ideal of ergodicity is inherently too constrained (see, e.g., Refs.\cite{heller_phys.today_46_38_2008, jensen_nature_355_311_1992, jensen_nature_355_591_1992, stone_phys.today_58_37_2005,einstein_verh.dtsch.phys.ges_19_82_1917, berry_phys.scripta_40_335_1989, Berry_rspa_413_182_1987}).

A standard way to assess the quantum degree of chaos (for more general discussion, see, e.g., Refs.~\cite{Heller_book, Gutzwiller_book, Haake_book, Tabor_book, Nakamura_book, Casati_book, Stockmann_book}) is through the statistical properties of its energy spectrum. This line of inquiry generally culminates in the Bohigas–Giannoni–Schmit (BGS) conjecture~\cite{Bohigas_phys.rev.lett_52_1_1984}, rooted in Random Matrix Theory~\cite{Mehta_book} (RMT): the spectral fluctuations of a quantum system whose classical counterpart is integrable follow Poisson statistics, while those of a classically chaotic system agree with the Gaussian orthogonal (GOE) or unitary (GUE) ensembles, depending on the presence or absence of time-reversal symmetry. Complementing this approach, system-specific structure is captured by periodic-orbit (PO) theory, most notably through the Gutzwiller trace formula~\cite{Gutzwiller_j.math.phys_12_343_1971}, which relates the quantum spectrum and density of states to classical periodic orbits. This connection further seems to suggest that no single PO should dominate in the contribution to any individual eigenstate, consistent with the spirit of the Berry conjecture~\cite{Berry_j.phys.a_10_2083_1977}: a typical eigenstate of a chaotic system is locally approximated by a random superposition of plane waves with the appropriate kinetic energy.

On the eigenstate side, rigorous quantum ergodicity theorems~\cite{Shnirelman_Uspekhi.Mat.Nauk_29_181_1974, Colindeverdiere_comm.math.phys_102_497_1985, zelditch_duke.math.j_55_919_1987} establish that the expectation value of any reasonable observable converges to its classical microcanonical average for almost all eigenstates, thereby supporting the eigenstate thermalization hypothesis (ETH)~\cite{Srednicki_phys.rev.a_50_888_1994, Deutsch_phys.rev.a_43_2046_1991}. Nevertheless, ETH has known exceptions, such as integrable systems~\cite{Rigol_phys.rev.lett_108_110601_2012} and many-body localized systems~\cite{abanin_rev.mod.phys_91_021001_2019, nandkishore_annu.rev.condens.matter.phys_6_15_2015}. In fact, even in a thermalizing system, not every eigenstate must satisfy ETH; the ergodicity theorems~\cite{Shnirelman_Uspekhi.Mat.Nauk_29_181_1974, Colindeverdiere_comm.math.phys_102_497_1985, zelditch_duke.math.j_55_919_1987} state that while a certain type of ergodization is guaranteed only in an asymptotic sense, it may unfold extremely slowly~\cite{kaplan_physica.d_121_1_1998}, and the theorems even admit the existence of exceptional non-ergodic states, such as quantum scars~\cite{Heller_phys.rev.lett_53_1515_1984, kaplan_ann.phys_264_171_1998,Kaplan_nonlinearity_12_R1_1999}. On the other hand, the stacking theorem recently established in Refs.~\cite{Antiscarring_1, Antiscarring_2} implies that a sufficiently wide energy window of eigenstates must collectively restore uniform coverage of phase space once the energy window exceeds the scale set by the shortest PO. Consequently, in the presence of scarring, there must exist complementary anti-scarred states, whose probability density is suppressed along the scar-forming orbit, to maintain the required phase-space uniformity, as demonstrated in the cases of a disordered quantum well~\cite{Antiscarring_1} and a spinor condensate~\cite{Antiscarring_2}.

The eigenstate and spectral diagnostics summarized above critically overlook the dynamical constraints on (quantum) phase-space distributions (see, e.g., Ref.~\cite{heller_phys.rev.a_35_1360_1987}). This omission is vital, as quantum temporal correlations are known to exceed bounds imposed by classical dynamics. However, there exist quantum probes in the time domain that more directly parallel classical notions of ergodic flow. A natural choice is to follow the time evolution of a localized wavepacket in coordinate or phase space.~\cite{Heller_book} Classically, the associated flux may linger near its phase-space origin, but in an ergodic system long-time averages show no sustained preference for the initial neighborhood~\cite{Arnold_book}. In contrast, quantum evolution violates this expectation dramatically. As shown in Ref.~\cite{quantum_birthmarks}, any type of quantum evolution entails a pervasive memory effect, termed as the quantum birthmark (QB). This phenomenon highlights a simple yet profound message: quantum dynamics is remain indelibly shaped by their initial conditions and early-time behavior, undermining any naive quantum analogues of the thermalization hypothesis.

More precisely, the QB is a permanent non-ergodic memory retained by a non-stationary state $\lvert a\rangle$ and all its time-evolved states $\lvert a(t)\rangle\equiv\lvert \alpha\rangle$. It manifests as an enhancement comparison to the classical expectation in the long-time average probability of occupying a state $\lvert b\rangle$ that is \emph{a priori} matched to the chosen initial state $\lvert a\rangle$, mathematically given by
\begin{equation}\label{Eq:fundamental_formula}
   \bar{P}_{ab} = \lim_{T \rightarrow \infty} \frac{1}{T} \int_0^T \vert \langle b\vert \alpha \rangle \vert^2 \, dt = 
   \sum\limits_n p_n^{a} p_n^{b},
\end{equation}
where $p_n^{a}=\lvert a_n\rvert^2$ and $p_n^{b}=\lvert b_n\rvert^2$ are the weights of the eigenstate $\lvert E_n\rangle$ in $\lvert a\rangle$ and $\lvert b\rangle$. As shown in Ref.~\cite{quantum_birthmarks}, the QB enhances the self-overlap $\bar{P}_{aa}$ relative to the overlap with a “typical’’ ergodic state $\lvert b\rangle\neq \lvert \alpha\rangle$ in the sense of
\begin{equation*}
    \frac{\bar{P}_{aa}}{\bar{P}_{ab}} \simeq P^{\textrm{UQB}} \cdot P^{\textrm{RQB}} > 2,
\end{equation*}
where the ratio can be effectively decomposed into the \emph{universal} and $P^{\mathrm{UQB}}$ and \emph{revival} factor $P^{\mathrm{RQB}}$.

The ratio $\bar{P}_{aa}/\bar{P}_{ab}$ is also raised above the ergodic standard (classical expectation) of unity by the universal factor of $P^{\textrm{UQB}} \ge 2$ depending on the global symmetries of the system and the revival enhancement factor of $P^{\textrm{RQB}} \ge 1$ encoding the effect of the early dynamics. In other words, every initial quantum state $\vert a \rangle$, along with the states $\vert \alpha \rangle$ it evolves into, experiences a universal enhancement factor of $P^{\textrm{UQB}}$ arising even if the spectrum has qualities expected of a random matrix (see, e.g., Ref.~\cite{kaplan_j.phys.A_40_F1063_2007}); whereas the revival factor $P^{\textrm{RQB}}$ takes into account short-time dynamics, correcting the RMT behavior for recurrences which might be due to a PO for example (see, e.g., Refs.~\cite{Smith_phys.rev.e_80_035205_2009, Smith_phys.rev.e_82_016214_2010}). For instance, this concept of revival-enchancement QB also seems to provide a unified framework for the various types of scarring, i.e., the Heller-type, variational and many-body scars.

In complementary to the work of Ref.~\cite{quantum_birthmarks}, we here develop the full theoretical framework for the universal QB. We provide a rigorous proof for $P^{\textrm{UQB}} \ge 2$ in fully chaotic quantum systems described by random matrix theory, and then extend this result to account for (possible) additional symmetries. We illustrate the UQB effect with numerical simulations of a billiard system, and also discuss the interpretation of quantum ergodicity modified by the omnipresent QQB.

\section{Universal birthmark}

Even in the absence of early-time revivals, i.e., for $P^{\textrm{RQB}}=1$, the measure defined in Eq.~\ref{Eq:fundamental_formula} already predicts a substantial memory effect arising purely from quantum interference. Specifically, the probability that the system returns to its initial state $\lvert a\rangle$ is given by
\begin{equation}\label{Eq:dilation_and_PR}
    \bar{P}_{aa} = \lim_{T \rightarrow \infty} \frac{1}{T} \int_0^T \vert \langle a \vert \mathcal{U}(t) \vert a \rangle \vert^2 \, dt = \sum_n \vert a_n \vert^4,
\end{equation}
which coincides with the inverse participation ratio, sometimes referred to as the dilation~\cite{Heller_book}. Notably, under unitary time evolution this quantity does not distinguish between regular and chaotic dynamics, or—within the framework of the BGS conjecture~\cite{Bohigas_phys.rev.lett_52_1_1984}—between integrable and chaotic spectra, since it carries no explicit dependence on spectral correlations. Instead, the value of $\bar{P}_{aa}$ is determined entirely by the initial state, namely by the coefficients $a_n$, in sharp contrast to the behavior expected from classical dynamics.

We further see that the factor $\bar{P}_{aa}$ is bounded from below by a purely quantum constraint,
\begin{equation}
    \bar{P}_{aa} - \frac{1}{N} = \sum_n \left( p_n^{a}  - \frac{1}{N}\right)^2 \ge 0 \, \Rightarrow \, \bar{P}_{aa} \ge \frac{1}{N}
\end{equation}
that follows directly from the normalization condition$\sum_n p_n^{a} = 1$. Consequently, quantum dynamics cannot yield the fully ergodic value $\bar{P}_{aa}=1/N$ unless the initial state is artificially chosen to be Haar-random-like, with $a_n = 1/\sqrt{N}$. We therefore argue that the initial configuration $\lvert a\rangle$ itself constitutes the nontrivial ingredient responsible for violating naive expectations of full ergodicity. Additionally, we can generally conclude
\begin{equation}
    \bar{P}_{aa} = \sum_n p_n^{a}p_n^{a} \ge \sum_n p_n^{a}p_n^{b} = \bar{P}_{ab},
\end{equation}
where $\vert b\rangle$ is dynamically distinguishable but otherwise {\it a priori}  matched to the initial state $\vert a\rangle$. The equality only holds for the initial state $\vert b\rangle = \vert a \rangle$ or any evolute  $\vert b  \rangle  = \vert \alpha \rangle$. This result stems from the fact that fluctuations in $p_n^{a}$ and $p_n^{b}$ from one $n$ to the next are independent, the squares of Gaussian independent random variables; whereas the state $\vert a \rangle$ obviously correlates with itself. These two observations underlie the ubiquitous appearance of the universal factor $P^{\textrm{UPQ}}$ in quantum systems.

In particular, we show 
\begin{equation}\label{Eq:universal_QB}
    P^{\textrm{UQB}} \sim 
    \begin{cases}
        2 & \textrm{without time-reversal symmetry}\\
        \\
        3 & \textrm{with time-reversal symmetry}\\
    \end{cases}.
\end{equation}
We furthermore show how this generalizes to the case of additional symmetries

\subsection{Full random matrix case}

We begin by considering a chaotic quantum system, meaning that its Hamiltonian commutes only with itself and the identity. Under this assumption, spectral degeneracies are absent. The enhancement factor $P^{\textrm{UQB}}\sim 2$ predicted by Eq.~\ref{Eq:universal_QB} for a generic, non-stationary state in such a system does not originate from system-specific dynamical features, but instead reflects fundamental statistical correlations, captured by RMT. 

A Haar-random quantum state $\lvert a\rangle$ has expansion amplitudes that are approximately independent Gaussian variables, coupled only through the normalization constraint $\sum_j |a_j|^2 = 1$. Albeit weak, this global constraint induces correlations among all components. Upon enforcing normalization, the weights $p_j = |a_j|^2$ no longer follow independent distributions but instead obey the Dirichlet distribution, $(p_1,\dots,p_N) \sim \mathrm{Dir}(\alpha,\dots,\alpha)$, with the concentration parameter $\alpha=1$ for the GUE and $\alpha=\tfrac{1}{2}$ for the GOE. As a consequence, the typical fourth moment $\langle |a_j|^4\rangle$ exceeds $\langle |a_j|^2\rangle^2$ by an amount that depends only on the Hilbert-space dimension $N$. This produces a universal excess of return probability relative to the fully random expectation $P_{ab} = 1/N$, or more formally
\begin{equation}\label{Eq:universal_QB_formula}
    \frac{\bar{P}_{aa}}{\bar{P}_{ab}} = 
    \begin{cases}
        \frac{2N}{1 + N} \sim  2\, (N \gg 1) & \textrm{for GUE}\\
        \\
        \frac{3N}{2 + N} \sim 3\, (N \gg 2) & \textrm{for GOE}\\
    \end{cases}.
\end{equation}
In the limit of large $N$, this expression reduces to the universal factor quoted in Eq.~\ref{Eq:universal_QB}.

The proof of this enhancement factor, presented in Appx.~\ref{Appendix:derivation}, proceeds in three conceptual steps. First, we characterize the statistics of the expansion coefficients of a Haar-random state (see Appx.~\ref{Sec:dirichlet_laws}): prior to normalization the amplitudes are Gaussian, while imposing $\sum_j |c_j|^2=1$ leads the squared moduli $p_j=|c_j|^2$ to follow a Dirichlet distribution, with parameters determined by the symmetry class (complex for GUE, real for GOE). Second, we compute the relevant second and fourth moments of this distribution, yielding $\mathbb{E}[p_i^2]$ and $\mathbb{E}[p_i p_j]$, and thus the averaged self-overlap $\bar P_{aa}$. An equivalent derivation based on Haar invariance and Schur–Weyl duality is also provided in Appx.~\ref{Sec:schur_weyl_argumentation}
(see also Ref.~\cite{quantum_birthmarks}); there, symmetry restricts the fourth-order moment tensor to a small set of invariant contractions fixed by normalization.
 
An alternative route to the same result relies on a well-established property of random matrix theory~\cite{Mehta_book, ORourke_j.comb.theory_144_361_2016}: the eigenvectors of GOE and GUE matrices are uniformly distributed on the unit sphere and independent of the eigenvalues, as a direct consequence of the invariance properties of the ensembles. The corresponding spheres are $S^{N-1}$ for GOE and $S^{2N-1}$ for GUE, reflecting the fact that complex amplitudes (GUE) carry twice as many degrees of freedom as real ones (GOE). Averaging the fourth moment of a generic state over these spheres directly reproduces Eq.~\ref{Eq:universal_QB_formula}.

\subsection{Additional symmetries}

The presence of additional symmetries calls for a more nuanced treatment. When the state is entirely confined to a single symmetry subspace, the familiar RMT enhancement factor of 2 or 3 is recovered. The same result holds if the state uniformly populates all symmetry-related subspaces. In the intermediate situation, however, when the state occupies these subspaces unevenly, the enhancement defined relative to the full Hilbert-space dimension typically exceeds the standard RMT value. By contrast, analyzing each symmetry subspace separately restores the usual RMT enhancement, fully consistent with our notion of a priori matching introduced above.

If, instead, the enhancement factor is defined with respect to the full effective Hilbert space (see Appx.~\ref{Appx:additional_symmetries}), the QB effect may appear artificially amplified. At the theoretical level, their influence can be absorbed into the definition of the universal factor $P^{\textrm{UQB}}$: once all relevant a priori constraints—both symmetry and energy—are properly accounted for, the symmetry-reduced description invariably yields the RMT bounds $P^{\textrm{UQB}}=2$ or $3$. Without symmetry reduction, however, additional symmetries generally produce an apparent enhancement exceeding the RMT expectation, i.e., $P^{\textrm{UQB}}>3$.


\section{Conclusion}

In summary, we have presented a complete theory for the UQB, a persistent memory effect intrinsic to unitary quantum  dynamics of a generic non-stationary state. Our findings reveal a more intricate and nuanced structure of non-ergodic behavior in classically chaotic systems than previously recognized. The notion of a QB transcends the ergodicity of individual eigenstates, as well as reestablishing a connection to the classical, dynamics-based standpoint of ergodicity. In this light, QBs offer an fresh but important perspective on the quantum nature of ergodicity.

\section*{Acknowledgments}

The authors thanks L. Kaplan and R. Ketzmerick for many inspiring discussions. This work was supported by the National Science Foundation under Grant No. 2403491. A.M.G. thanks the Studienstiftung des Deutschen Volkes for financial support. J.K.-R. thanks the Oskar Huttunen, Vaisala, Emil Aaltonen and Magnus Ehrnrooth Foundations for the financial support.

\appendix

\section{Random matrix value}\label{Appendix:derivation}

Consider two generic states $\ket{a}$ and $\ket{b}$ in an $N$-dimensional Hilbert space $\mathcal{H}$, spanned by an eigenstate basis $\ket{E_i}$ $(i=1,\cdots,N)$ of the given Hamiltonian $H$ with corresponding eigenenergies $E_i$. Assume for simplicity that the spectrum of the Hamiltonian is non-degenerate. Then the state $\ket{a}$ can be decomposed as
\[
\ket{a} = \sum_{i=1}^{N} c_j^{(a)} \ket{E_i}, \text{ with } c_i^{(a)} = \braket{a | E_i}.
\]

Similarly we can decompose $\ket{b}$ as $\ket{b} = \sum_{i} c_i^{(b)} \ket{E_i}$. Write
\[
p_j \equiv |c_j|^2, \text{ with the normalization constraint }\sum_{j=1}^N p_j = 1.
\]

We are interested in moments of the coordinates $p_j$, notably $\mathbb{E}[p_i^2]$ and the cross–moment $\mathbb{E}[p_i p_j]$ with $i\neq j$, and in the long-time averages
\[
\bar{P}_{aa} \equiv \mathbb{E}\!\left[\sum_{j=1}^N p_j^2\right], 
\qquad 
\bar{P}_{ab} \equiv \mathbb{E}\!\left[\sum_{j=1}^N |c_j^{(a)}|^2 |c_j^{(b)}|^2\right],
\]
where $\ket{a}$ and $\ket{b}$ are independent random states unless otherwise stated.  

The return (self–overlap) average is
\[
\bar{P}_{aa}=\mathbb{E}\!\left[\sum_{j=1}^N p_j^2\right]=
\begin{cases}
N \cdot \dfrac{2}{N(N+1)} = \dfrac{2}{N+1}, & \text{GUE},\\[8pt]
N \cdot \dfrac{3}{N(N+2)} =\dfrac{3}{N+2}, & \text{GOE},
\end{cases}
\]
and for two independent random states $\ket{a},\ket{b}$ one finds
\[
\bar{P}_{ab}=\mathbb{E}\!\left[\sum_{j=1}^N |c_j^{(a)}|^2 |c_j^{(b)}|^2\right] = \frac{1}{N}
\],
as we will prove below. Hence
\[
\frac{\bar{P}_{aa}}{\bar{P}_{ab}}=
\begin{cases}
\dfrac{2N}{N+1}, & \text{GUE},\\\\

\dfrac{3N}{N+2}, & \text{GOE}.
\end{cases}
\]

\subsection{Distribution of coordinates: Dirichlet laws}\label{Sec:dirichlet_laws}

Starting with i.i.d.\ Gaussian coordinates:
\begin{itemize}
\item GUE (complex): $c_j = x_j + i y_j$ with $x_j,y_j \sim \mathcal{N}(0,\tfrac{1}{2N})$ so that $|c_j|^2 \sim \Gamma(1,\tfrac{1}{N})$.
\item GOE (real): $c_j \sim \mathcal{N}(0,\tfrac{1}{N})$ so that $c_j^2 \sim \Gamma(\tfrac{1}{2}, \tfrac{2}{N})$.
\end{itemize}
Normalizing by the $\ell^2$ norm induces the probability projection vector $p=(p_1,\dots,p_N)$, which obeys a Dirichlet distribution:
\[
(p_1,\dots,p_N) \sim 
\begin{cases}
\mathrm{Dir}(1,\dots,1) & \text{GUE},\\[2pt]
\mathrm{Dir}\!\left(\tfrac{1}{2},\dots,\tfrac{1}{2}\right) & \text{GOE}.
\end{cases}
\]

\paragraph{Consequences.}
\begin{itemize}
\item GUE: $\alpha_i=1$, $\alpha_0=N$,
\[
\;\mathbb{E}[p_i^2]=\frac{2}{N(N+1)},\quad \mathbb{E}[p_i p_j]=\frac{1}{N(N+1)}\;(i\ne j). \;
\]
\item GOE: $\alpha_i=\tfrac{1}{2}$, $\alpha_0=\tfrac{N}{2}$,
\[
\;\mathbb{E}[p_i^2]=\frac{3}{N(N+2)},\quad \mathbb{E}[p_i p_j]=\frac{1}{N(N+2)}\;(i\ne j). \;
\]
\end{itemize}
Both agree asymptotically with the i.i.d.\ Gaussian heuristic as $N\to\infty$.

\subsection{Schur–Weyl argument}\label{Sec:schur_weyl_argumentation}

Define the fourth–order moment tensors
\begin{align}
T_{\alpha\beta\gamma\delta} \equiv \mathbb{E}\!\left[ c_\alpha c_\beta^* c_\gamma c_\delta^* \right]\quad\text{(GUE)},\\ 
T_{ijkl} \equiv \mathbb{E}\!\left[ c_i c_j c_k c_l \right]\quad\text{(GOE)}.
\end{align}
Haar (unitary/orthogonal) invariance implies that $T$ lives in the commutant of the group action. By Schur–Weyl duality, the invariant tensors are linear combinations of index pairings:
\begin{align}
\text{GUE: }~ T_{\alpha\beta\gamma\delta} &= A\,\delta_{\alpha\beta}\delta_{\gamma\delta} + B\,\delta_{\alpha\delta}\delta_{\beta\gamma},\\
\text{GOE: }~ T_{ijkl} &= X\,\delta_{ij}\delta_{kl}+Y\,\delta_{ik}\delta_{jl}+Z\,\delta_{il}\delta_{jk}.
\end{align}
Let $C\equiv \mathbb{E}[|c_j|^4]$ and $D\equiv \mathbb{E}[|c_i|^2|c_j|^2]$ for $i\ne j$ (GUE), and similarly $C\equiv \mathbb{E}[c_j^4]$, $D\equiv \mathbb{E}[c_i^2 c_j^2]$ (GOE). The normalization $\sum_j |c_j|^2=1$ gives
\[
N C + N(N-1) D = 1.
\]
Matching index patterns yields: in GUE, $A=B=D$ and $C=2D$; in GOE, $X=Y=Z=D$ and $C=3D$. Solving gives
\begin{align}
\text{GUE: } C&=\frac{2}{N(N+1)} \quad  D=\frac{1}{N(N+1)}\\ 
\text{GOE: } C&=\frac{3}{N(N+2)} \quad D=\frac{1}{N(N+2)}.
\end{align}
These agree with the Dirichlet evaluation above.

\subsection{Symmetries and sector restrictions}\label{Appx:additional_symmetries}

Let a symmetry group $G$ act unitarily on $\mathcal{H}$ with a multiplicity–free decomposition
\[
\mathcal{H}=\bigoplus_{\alpha} \mathcal{H}_\alpha,\qquad d_\alpha=\dim \mathcal{H}_\alpha,\qquad \sum_\alpha d_\alpha=N.
\]
Write the projection of $\ket{a}$ to sector $\alpha$ as $c_\alpha \ket{a_\alpha}$ with $\sum_\alpha |c_\alpha|^2=1$. If \emph{both} $\ket{a}$ and $\ket{b}$ are random on the full unit sphere (i.e.\ no restriction to a subset of sectors), then averaging over the Haar measure yields the same $\bar{P}_{aa}$ and $\bar{P}_{ab}$ as above, independent of symmetry: the enhancement remains 
\[
\eta=\frac{2N}{N+1}\;\;(\text{GUE}),\qquad \eta=\frac{3N}{N+2}\;\;(\text{GOE}).
\]
To obtain symmetry \emph{amplification}, the state(s) must be restricted. Suppose $\ket{a}$ is supported only inside a $G$–invariant subspace of total dimension $d=\sum_{\alpha' } d_{\alpha'}$ (sum over accessible sectors), while $\ket{b}$ is still generic on $\mathcal{H}$. Repeating the Dirichlet/Schur–Weyl calculation inside the $d$–dimensional accessible space gives
\[
\bar{P}_{aa}=
\begin{cases}
\dfrac{2}{d+1}, & \text{GUE},\\[4pt]
\dfrac{3}{d+2}, & \text{GOE},
\end{cases}
\qquad
\bar{P}_{ab}=\dfrac{1}{N},
\]
and therefore
\[
\frac{\bar{P}_{aa}}{\bar{P}_{ab}}=
\begin{cases}
\dfrac{2N}{d+1}, & \text{GUE},\\[6pt]
\dfrac{3N}{d+2}, & \text{GOE}.
\end{cases}
\]
These reduce to the universal (unrestricted) formulas when $d=N$.

\bibliography{references}

\end{document}